\documentclass[amsmath,amssymb,preprint,nofootinbib,floatfix,showpacs,showkeys]{revtex4}
\usepackage[dvipdfm]{graphicx}
\usepackage[dvipdfm]{hyperref}
\usepackage{multirow}
\newcommand{\bo}{\raise-0.4mm\hbox{$\Box$}} 

\begin{document}

\title{Effective Potential for Complex Langevin Equations}

\author{Gerald Guralnik}
\email{gerry@het.brown.edu}
\author{Cengiz Pehlevan}
\email{cengiz@het.brown.edu}
\affiliation{Department of Physics, Brown University, Box 1843, Providence, RI 02912}
\date{\today}

\begin{abstract}  
  We construct an effective potential for the complex Langevin equation on a
  lattice. We show that the minimum of this effective potential gives the
  space-time and Langevin time average of the complex Langevin field. The loop
  expansion of the effective potential is matched with the derivative
  expansion of the associated Schwinger-Dyson equation to predict the
  stationary distribution to which the complex Langevin equation converges.
\end{abstract}


\preprint{Brown-HET-1571}
\maketitle

\tableofcontents

\section{Introduction}

Sampling complex integral weights of the form $e^{-S}$ is important for many
applications in high energy physics. Complex Langevin equations have been proposed
for this purpose \cite{Parisi83,Klauder83}. Many theoretical questions are
still to be answered for complex Langevin equations. Most importantly, it is
not known a priori if a complex Langevin simulation will converge to a
stationary distribution. Even if it does, there are multiple stationary
distributions that satisfy stationary limit equations (Fokker-Planck equation
\cite{Salcedo93} or Schwinger-Dyson equation \cite{Pehlevan07}) and it is not
known how to identify to which one of these that the complex Langevin equation
converges. See \cite{Pehlevan07} for a discussion of these issues and
references to literature.

Identification of the particular stationary distribution sampled by the
complex Langevin equation needs input from the dynamics of the complex
Langevin simulation. Analysis depending only on the quantities in the
stationary limit has not been successful in differentiating between stationary
distributions. To account for this fact, we introduce an effective potential
approach to complex Langevin equations. Even though this effective potential
does not have all the properties of the conventional quantum field theory
effective action, we show that it governs the probability that the spacetime
and Langevin time average of the fields takes on specific values. In the
infinite Langevin time limit, it is shown that the spacetime and Langevin time
average of the field must be equal to the minimum of the effective potential. As
in quantum field theory, a loop expansion for this effective action can
be defined. The parameter that defines loops is a factor that multiplies the
noise in the complex Langevin equation. The same parameter is shown to count
derivatives in the Schwinger-Dyson equation derived from the action $S$. This
point is important, because it is known that the stationary distributions of
the complex Langevin equation are complexified path integral solutions of the
Schwinger-Dyson equation, i.e. see \cite{Pehlevan07}. The main idea of this
paper is to use the minimum of the effective action to identify the stationary
distribution to which the complex Langevin equation converges by matching the
loop expansion with the derivative expansion. This provides an approach to
addressing the multiple stationary distributions problem.

In two recent studies \cite{Aarts08,Berges0708}, the classical fixed point
structure of complex Langevin equations were shown to be important for the
convergence behavior. Our results will also parallel this result since to leading
order the minima of the effective potential will be given by classical fixed
points. The next order will be used to remove the degeneracy and this will lead
us to the identification of the stationary distribution that is sampled by the
complex Langevin simulation.

We start by reviewing relevant results in Section \ref{one}. We then introduce
the effective potential in a lattice setting, discuss its properties and
calculate it to one loop order (Section \ref{two}). Before we conclude, we
give a simple application of these ideas (Section \ref{Examples}).

One point we omit in this paper is the discussion of the continuum limit. The
question of renormalizability is a nontrivial problem and deserves a separate
study of its own.

\section{Stationary Distributions of Complex Langevin Equations}\label{one}

In this section we review the relevant results of \cite{Pehlevan07}. Even
though, our ultimate goal in this paper is to do calculations in a lattice
setting, the arguments in this section are given in the continuum for ease of
notation. The lattice versions of these results are either in
\cite{Pehlevan07} or \cite{Berges07}.

Given a complex action $S$ of a scalar field $\phi(x)$ in a Euclidean space, 
\begin{equation}
  S[\phi] = \int d^dx\, \left[\frac 12\phi(x)\left(-\Box + m^2\right)\phi(x) + \sum_l \frac{g_l}{l}\phi^l(x)\right],
\end{equation}
we want to study the stochastic dynamics given by
\begin{equation}\label{modified}
\frac{\partial\phi(x,\tau)}{\partial\tau} = -\frac{\delta S[\phi]}{\delta\phi(x,\tau)} + \eta^R(x,\tau) + i\eta^I(x,\tau),
\end{equation}
where $\eta^R$ and $\eta^I$ are independent Gaussian noises with correlators given by
\begin{align}\label{corr}
\left\langle\eta^R(x,\tau)\right\rangle &= 0, \nonumber \\
\left\langle\eta^R(x,\tau)\,\eta^R(x',\tau')\right\rangle &= 2 \Omega\alpha\delta^d(x-x')\,\delta(\tau-\tau'),
\end{align}
and
\begin{align}\label{corr1}
\left\langle\eta^I(x,\tau)\right\rangle &= 0, \nonumber \\
\left\langle\eta^I(x,\tau)\,\eta^I(x',\tau')\right\rangle &= 2 \Omega\left(\alpha-1\right)\delta^d(x-x')\,\delta(\tau-\tau').
\end{align}
$\alpha$ and $\Omega$ are real constants and $\alpha\geq 1$. $\Omega=1$ corresponds to the complex Langevin equation. We will not set $\Omega$ to any particular value for the moment. $\Omega$ will define orders in the approximations we do below. Equation \eqref{modified} can be written as a stochastic system of two real fields, $\phi(x,\tau) = \phi^R(x,\tau) + i\phi^I(x,\tau)$. In fact, this is what we will do in the rest of this paper.

Note that most of the time in the literature $\alpha$ is set to 1. There is only a real noise in the complex Langevin equation. As shown in \cite{Pehlevan07}, this may lead to initial condition dependence of the stationary distribution sampled by the complex Langevin equation. This happens because for some actions the complex Langevin dynamics may be confined to sampling in lower or upper half planes depending on the initial condition, not having access to all of the complex plane. In this paper, we want to avoid such initial condition dependence and assume that the stationary distribution sampled by the complex Langevin equation is unique. Furthermore, setting $\alpha$ to some value larger then 1 will be important for deriving a path integral representation of the generating functional for the complex Langevin process in the next section.

Given some initial conditions for the field, Langevin equation \eqref{modified} has a time dependent probability distribution, see e.g \cite{Justin02},
\begin{equation}\label{timeprob}
P\left(\tau,\phi^R(x),\phi^I(x)\right) = \left\langle\delta\left(\phi^R(x,\tau)-\phi^R(x)\right)\,\delta\left(\phi^I(x,\tau)-\phi^I(x)\right)\right\rangle_{\eta^R,\eta^I},
\end{equation}
which can be shown to satisfy a Fokker-Planck equation,
\begin{align}\label{FP}
 \frac{\partial P\left(\tau,\phi^R,\phi^I\right)}{\partial \tau} = & \int d^dx \left[\frac{\delta}{\delta\phi^R(x)}\left(\Omega\alpha\frac{\delta P}{\delta \phi^R(x)}+\text{Re}{\frac{\delta S[\phi]}{\delta \phi(x)}}P\right) \right.  \nonumber \\
&\qquad\qquad+\left.\frac{\delta}{\delta\phi^I(x)}\left(\Omega\left(\alpha-1\right)\frac{\delta P}{\delta \phi^I(x)}+\text{Im}{\frac{\delta S[\phi]}{\delta \phi(x)}}P\right)\right].
\end{align}
The existence and uniqueness of stationary distributions for this Fokker-Planck equation is still an unanswered question. 

We continue by defining a generating functional for equal time correlation functions
\begin{equation}\label{equaltime}
  Z[J(x),\tau) \equiv \left\langle e^{\int d^dx \left(\phi^{R,\eta}(x,\tau)+i\phi^{I,\eta}(x,\tau)\right)J(x)}\right\rangle_{\eta^R,\eta^I},
\end{equation}
where $J(x)$ is a real source field. Here the brackets denote an average over the noise. Field variables are considered as functionals of the noise through equation \eqref{modified}, hence we include a superscript $\eta$. The effect of noise appears in the stochastic average through the probability distribution $P$ (see e.g. \cite{Justin02}),
\begin{equation}
Z[J,\tau) = \int [d\phi^R][d\phi^I]\,P(\tau,\phi^R,\phi^I) \, e^{\int d^dx \left(\phi^R(x,\tau)+i\phi^I(x,\tau)\right)J(x)}.
\end{equation}
This is now an integral over all field configurations weighted by $P$. This point will be treated in more detail in the next section. Using the Fokker-Planck equation \eqref{FP}, one can show that the following equation holds
\begin{align}\label{Ztau}
  \frac {\partial Z[J,\tau)}{\partial \tau} = \int d^dx \,\left\lbrace J(x)\left[\Omega J(x)-\frac{\delta S[\phi]}{\delta \phi(x)}\left(\frac{\delta}{\delta J}\right)\right]\right\rbrace Z[J,\tau).
\end{align}
If there is a stationary distribution, the left hand side is zero. When
$\Omega=1$, functional derivatives of the right hand side with respect to $J$
produces Schwinger-Dyson identities for the action $S[\phi]$, since the
integrand must vanish due to the arbitrariness of $J$. $\Omega$ plays the role
of $\hbar$ in quantum field theory. It is easy to see this by defining a
making a change of variables $J(x) \rightarrow J(x)/\Omega$.\footnote{In terms
  of this new variable, the correlation functions are given by functional
  derivatives of $Z[J,\tau)$ with respect to $\Omega\frac{\delta}{\delta J}$.}
Then, the stationary distribution condition becomes
\begin{equation}
\int d^dx \,\left\lbrace \frac{J(x)}{\Omega}\left[J(x)-\frac{\delta S[\phi]}{\delta \phi(x)}\left(\Omega\frac{\delta}{\delta J}\right)\right]\right\rbrace Z[J,\tau)=0.
\end{equation}
$\Omega$ counts the number of functional derivatives, as $\hbar$ would. As is
well known, see \cite{Bender77}, a derivative expansion of Schwinger-Dyson
equations is equivalent to a loop expansion of the associated path integral
around a local minimum.

If one considers zero dimensional theories, an explicit construction of the stationary equal time correlation functional can be given. Generalization to a lattice is straightforward, see \cite{Pehlevan07}.  Now, the action $S$ is a function of $\phi$. Equation \eqref{Ztau} becomes
\begin{equation}
\frac {\partial Z(j,\tau)}{\partial \tau} = \Omega j^2Z(j,\tau) - j\frac{\partial S}{\partial \phi} \left(\frac{\partial}{\partial j}\right)Z(j,\tau).
\end{equation}
In particular, the generating function for the stationary distribution satisfies
\begin{equation}\label{SD}
 \frac{\partial S}{\partial \phi} \left(\frac{d}{dj}\right)Z_{st}(j)=\Omega jZ_{st}(j).
\end{equation}
We can solve this equation following \cite{Garcia96}. Our ansatz is  
\begin{equation}\label{ansatz}
  Z_\Gamma(j) = \int_\Gamma d\phi \, G(\phi) e^{j\phi},
\end{equation}
where $\Gamma$ is a contour over the complex plane. Inserting this into equation (\ref{SD}) one gets:
\begin{equation}
0 = - \left. \Omega G(\phi) e^{j\phi}\right|_{\partial\Gamma} + \int_\Gamma d\phi \left[\frac{\partial S}{\partial \phi}G(\phi) + \Omega \frac{dG(\phi)}{d\phi}\right]e^{j\phi}. 
\end{equation}
This equation can be solved for 
\begin{equation}
G(\phi) = e^{-\frac{S(\phi)}{\Omega}},
\end{equation}
and $\Gamma$ is contour that connects the zeros of $e^{-S(\phi)+j\phi}$ on the complex $\phi$ plane. As advertised, $\Omega$ plays the role of $\hbar$. Since equation \eqref{SD} is linear, any linear combination of the (independent) solutions will also be a solution, 
\begin{equation}\label{st}
  Z_{st}(j) = \sum_{\Gamma_I}a_{\Gamma_I} Z_{\Gamma_I}(j),
\end{equation}
where $\Gamma_I$ define an independent subset of paths $\Gamma$.\footnote{Essentially the same result is derived by another method in \cite{Salcedo93}.} The only constraint on the coefficients at this point is the normalization condition for correct distribution interpretation, namely that $Z_{st}(0)=1$. We restricted the form of the generating function, but did not identify it completely. Examples are given in \cite{Pehlevan07}. 

In this paper, we use an effective action approach to identify the stationary distribution. The $\Omega$ expansion gives a small noise expansion for the complex Langevin equation and a loop expansion of the stationary distribution generating function. Matching of both series order by order will tell us the exact form of the stationary distribution. 

\section{Effective Action for Complex Langevin Equation}\label{two}

In this section, we will define the effective action for complex Langevin equation and do an $\Omega$ expansion to gather information about the stationary distribution. We choose to use a lattice regularization. We work with an Ito type discretization for equation \eqref{modified}. 

First we define our notation. We introduce a Euclidean spacetime lattice with isotropic lattice spacing $a$. Lattice points are denoted by a $d$ dimensional vector $n=(n_1,\ldots,n_d)$, where $n_i=0,\ldots,(N-1)$. The coordinate of a lattice point is given by $x_i = n_ia$.  Our convention is to denote Euclidean time with $x_d$. Periodic boundary conditions are assumed for the space-time lattice. We discretize Langevin time $\tau$ with $\Delta\tau$ intervals. Langevin simulation is assumed to start at $\tau = \tau_i$ and run for $N_\tau$ time steps untill $\tau_f=\tau_i+N_\tau\Delta\tau$. In terms of dimensionless lattice variables $\hat{\phi}_n(\kappa) = a^{(d-2)/2}\phi(na,\kappa\Delta\tau)$, $\hat{\epsilon} = \Delta\tau/a^2$, $\hat{m} = am$, $\hat{g}_l= a^{d(1-l/2)+l}g_l$, $\hat{\eta}^{R,I}= \sqrt{a^d\Delta\tau}\eta^{R,I}$, where the noise correlators are now
\begin{align}
\left\langle\hat{\eta}^R_n(\kappa)\,\hat{\eta}^R_{n'}(\kappa')\right\rangle &= 2 \Omega\alpha\delta_{nn'}\delta_{\kappa\kappa'}, \nonumber \\
\left\langle\hat{\eta}^I_n(\kappa)\,\hat{\eta}^I_{n'}(\kappa')\right\rangle &= 2 \Omega\left(\alpha-1\right)\delta_{nn'}\delta_{\kappa\kappa'},
\end{align}
the discretized equation \eqref{modified} in Ito calculus reads
\begin{equation}\label{discretized}
  \hat{\phi}_n(\kappa+1) = \hat{\phi}_n(\kappa) - \hat{\epsilon}\left(-\hat{\bo}+\hat{m}^2\right)\hat{\phi}_n(\kappa) - \hat{\epsilon}\sum_l\hat{g}_l\hat{\phi}_n^{l-1}(\kappa)+\sqrt{\hat{\epsilon}}\left(\hat{\eta}^R_n(\kappa) + i\hat{\eta}^I_n(\kappa)\right).
\end{equation}
In the rest of this paper we assume $\alpha >1$. To avoid confusions, we note that $\hat{\phi}_n(\kappa)$ is the lattice variable at the spacetime point $na$ and Langevin time $\tau_i+\kappa\Delta\tau$.Here, the dimensionless lattice Laplacian is defined by
\begin{equation}
  \hat{\bo}\hat{\phi}_n(\kappa) = \sum_m\sum_{k=1}^d \left(\delta_{m,n+\hat{e}_k} + \delta_{m,n-\hat{e}_k}-2\delta_{m,n}\right)\hat{\phi}_m(\kappa),
\end{equation}
where $\hat{e}_\mu$ is a unit vector pointing along the $k$-direction. We assume an initial condition is set at $\tau=\tau_i$ by specifying the field on the spacetime lattice.

\subsection{Generating Functional for Dynamic Correlation Functions}

First step in our anaylsis is to define a generating functional for dynamic correlation functions,
\begin{equation}
Z_d[\hat{J^R},\hat{J}^I] \equiv \left\langle \exp\left[ \sum_{\kappa=1}^{N_\tau}\sum_{n} \hat{J}_n^R(\kappa)\hat{\phi}^{R,\eta}_n(\kappa) + \hat{J}_n^I(\kappa)\hat{\phi}^{I,\eta}_n(\kappa)\right] \right\rangle_{\hat{\eta}^R,\hat{\eta}^I}.
\end{equation}
$\phi^\eta$ is a functional of the noise fields through the Langevin equation \eqref{discretized}. The sum in $\kappa$ parameter starts from $1$, since we assume an initial condition $\phi_n(0)$ is given. Unless explicitly stated, $\kappa$ sums and products run from $0$ to $N_\tau-1$. We use subscript $d$ to differentiate $Z_d$ from the generating functional of equal time correlation functions defined by equation \eqref{equaltime}. $Z_d$ is more general than $Z$ since it generates correlation functions of the field variable at different times as well as equal times. We will define an effective action using this generating functional. Before doing that, we derive a path integral representation for $Z_d$.  Although we will not follow their work directly, see \cite{Kawara90} for a discussion of path integral representations of generating functionals based on Ito discretization and \cite{Gozzi83} for continuum formulation. By definition, $Z_d$ is an average over the probability density functions of the Gaussian random variables, 
\begin{align}
Z_d[\hat{J^R},\hat{J}^I] = \mathcal{C}\int \prod_{n,\kappa}&d\hat{\eta}_n^R(\kappa) d\hat{\eta}_n^I(\kappa) \, \exp\left[-\sum_{n,\kappa}\left(\frac{\hat{\eta}_n^R(\kappa)^2}{4\Omega\alpha} + \frac{\hat{\eta}_n^I(\kappa)^2}{4\Omega(\alpha-1)}\right)\right] \nonumber \\
&\times \exp\left[ \sum_{n}\sum_{\kappa=1}^{N_\tau}\left(\hat{J}_n^R(\kappa)\phi^{R,\eta}_n(\kappa) + \hat{J}_n^I(\kappa)\phi^{I,\eta}_n(\kappa)\right)\right].
\end{align}
$\mathcal{C}$ is a proper normalization constant for each equation it appears
in. The next step is to rewrite this expression as an integral over field
configurations. Using equation \eqref{discretized}, we can change integration
variables from the Gaussian random variables $\hat{\eta}_n^R(\kappa)$ and
$\hat{\eta}^I_n(\kappa)$ to $\hat{\phi}_n^R(\kappa+1)$ and
$\hat{\phi}_n^I(\kappa+1)$ . Note that due to our discretization, the Jacobian
of this transformation is a constant. Defining
\begin{align}
\hat{\Delta}\hat{\phi}_n (\kappa) &= \sum_\gamma(\delta_{\gamma,\kappa+1}-\delta_{\gamma,\kappa})\hat{\phi}_n(\gamma), \nonumber \\
\hat{\mathcal{D}}\hat{\phi}_n(\kappa) &= \left(\hat{\Delta}-\hat{\epsilon}\hat{\bo}\right)\hat{\phi}_n (\kappa),\nonumber \\
\hat{F}^R_n(\kappa)&=\hat{F}^R\left(\hat\phi_n^R(\kappa),\hat\phi_n^I(\kappa)\right) = \text{Re}\left[\hat{m}^2 \hat{\phi}_n(\kappa)+ \sum_l\hat{g}_l\hat{\phi}_n^{l-1}(\kappa)\right], \nonumber \\
\hat{F}^I_n(\kappa)&=\hat{F}^I\left(\hat\phi_n^R(\kappa),\hat\phi_n^I(\kappa)\right) = \text{Im}\left[\hat{m}^2\hat{\phi}_n(\kappa) + \sum_l\hat{g}_l\hat{\phi}_n^{l-1}(\kappa)\right],
\end{align}
the generating functional $Z_d$ becomes
\begin{align}\label{Zd}
Z_d[\hat{J^R},\hat{J}^I] = &\mathcal{C} \int  [d\hat{\phi}] \,  \exp\left[ \sum_{n}\sum_{\kappa=1}^{N_\tau}\left(\hat{J}_n^R(\kappa)\hat{\phi}^{R}_n(\kappa) + \hat{J}_n^I(\kappa)\hat{\phi}^{I}_n(\kappa)\right)\right] \nonumber \\ 
&\,\times \exp\left\lbrace -\frac{1}{\Omega}\sum_{n,\kappa}\left[\frac{\left[\hat{\mathcal{D}}\hat{\phi}_n^R(\kappa) + \hat{\epsilon}\hat{F}^R_n(\kappa)\right]^2 }{4\alpha \hat{\epsilon}} + \frac{\left[\hat{\mathcal{D}}\hat{\phi}_n^I(\kappa) + \hat{\epsilon}\hat{F}^I_n(\kappa)\right]^2}{4(\alpha-1)\hat{\epsilon}}\right]\right\rbrace,
\end{align}
where the measure is
\begin{align}
 [d\hat{\phi}] = \prod_n\prod_{\kappa=1}^{N_\tau}d\hat{\phi}^R_n(\kappa)d\hat{\phi}^I_n(\kappa)
\end{align}
There is no integration over $\hat{\phi}_n(0)$, since we assume it to be set by initial conditions. The exponent in the second line, with the normalization factor $\mathcal{C}$, is the probability density of a particular realization of the Langevin process given some initial $\hat{\phi}_n(0)$. We will denote it by $T\left(\hat{\phi}(N_\tau)|\hat{\phi}(0)\right)$. The previously defined probability distribution \eqref{timeprob} and $T\left(\hat{\phi}(N_\tau)|\hat{\phi}(0)\right)$ are very closely related, see e.g. \cite{Namiki92}. For $T$, $1/\Omega$ is an overall factor for the exponent, which will allow us to define a loop expansion for this generating functional. 

For later convenience, we define the quantity in the exponent of the second line of equation \eqref{Zd} as 
\begin{align}\label{pathaction}
I[\hat{\phi}^R,\hat{\phi}^I]=\sum_{n,\kappa}\left[\frac{\left[\hat{\mathcal{D}}\hat{\phi}_n^R(\kappa) + \hat{\epsilon}\hat{F}^R_n(\kappa)\right]^2 }{4\alpha \hat{\epsilon}} + \frac{\left[\hat{\mathcal{D}}\hat{\phi}_n^I(\kappa) + \hat{\epsilon}\hat{F}^I_n(\kappa)\right]^2}{4(\alpha-1)\hat{\epsilon}}\right].
\end{align}
Forgetting the complex Langevin equation roots for one moment, one can think of equation \eqref{Zd} as a the generating functional of a field theory on a  $d+1$ dimensional lattice with an action given by equation \eqref{pathaction}. This is not a usual field theory, because it does not have translational invariance due to $\kappa=0$ terms appearing in equation \eqref{pathaction}. One end of the lattice ($\kappa=0$) is fixed, whereas the other end ($\kappa=N_\tau$) is free. This picture will be useful later when we discuss the effective potential.

Before continuing, we note that having $\alpha >1$ was a great convenience for this calculation. Otherwise, we would start with averages over only the real noise variables and this would lead to a difficulty when changing variables in the path integral from noise to field variables. The number of field variables $\hat{\phi}^{R,I}_n(\kappa)$ would be half of $\hat\eta^R(\kappa)$. Therefore, one would need to choose half of the field variables as independent ones and express the remaining in terms of the independent ones. Having $\alpha>1$ saved us from these difficulties.

\subsection{Effective Action, Effective Potential and Their Interpretation}

Now we do the standart quantum field theoretic procedure to define an effective action for this generating functional,
\begin{align}
W[\hat{J^R},\hat{J}^I]&= \ln Z_d[\hat{J^R},\hat{J}^I], \nonumber \\
\Gamma[\hat{\phi}^{R},\hat{\phi}^{I}] &= - W[\hat{J}^{R,\hat{\phi}},\hat{J}^{I,\hat{\phi}}] + \left[ \sum_{n}\sum_{\kappa=1}^{N_\tau}\left(\hat{J}_n^{R,\hat{\phi}}(\kappa)\hat{\phi}^{R}_n(\kappa) + \hat{J}_n^{I,\hat{\phi}}(\kappa)\hat{\phi}^{I}_n(\kappa)\right)\right],
\end{align}
where
\begin{align}
  \hat{\phi}^{R,J}_n(\kappa) = \frac{\partial W[\hat{J^R},\hat{J}^I]}{\partial \hat{J}^R_n(\kappa)}, &\qquad \hat{\phi}^{I,J}_n(\kappa) = \frac{\partial W[\hat{J^R},\hat{J}^I]}{\partial \hat{J}^R_n(\kappa)}, \nonumber \\
 \hat{\phi}^{R,J}_n(\kappa) =  \hat{\phi}^{R}_n(\kappa) \qquad &\text{if} \qquad \hat{J}_n^{R}(\kappa) = \hat{J}_n^{R,\hat{\phi}}(\kappa), \nonumber \\
 \hat{\phi}^{I,J}_n(\kappa) =  \hat{\phi}^{I}_n(\kappa) \qquad &\text{if} \qquad \hat{J}_n^{I}(\kappa) = \hat{J}_n^{I,\hat{\phi}}(\kappa).
\end{align}
By definition, it follows that
\begin{equation}
\frac{\partial \Gamma[\hat{\phi}^{R},\hat{\phi}^{I}]}{\partial \hat{\phi}^R_n(\kappa)} = \hat{J}_n^{R,\hat{\phi}}(\kappa), \qquad \frac{\partial \Gamma[\hat{\phi}^{R},\hat{\phi}^{I}]}{\partial \hat{\phi}^I_n(\kappa)} = \hat{J}_n^{I,\hat{\phi}}(\kappa).
\end{equation}
For field configurations that are homogenous and static in Langevin time, the effective action reduces to ``effective potential'',
\begin{equation}
  \mathcal{V}(\hat{\phi}^{R},\hat{\phi}^{I}) = \frac{\Gamma[\hat{\phi}^{R},\hat{\phi}^{I}]}{N^dN_\tau}.
\end{equation}
In the next section, we will calculate the effective potential to one loop order (first order in $\Omega$). 

We note that the effective action defined above does not have all the
properties one remembers from quantum field theory. In particular, it does not
have the energy interpretation \cite{Coleman88} because the complex Langevin
equation \eqref{modified} as a system of two real Langevin equations is not
derivable from an action.\footnote{A Langevin system that is derived from an
  action will have component equations $\dot{q_i} = -\frac{\partial
    S}{\partial q_i} + \eta_i$.} See \cite{Hochberg99} for a detailed
discussion. However, one useful property is still valid. The effective
potential governs the probability that the spacetime and Langevin time average
of the field takes specific values. To show this, we follow closely the
argument of \cite{Hochberg99}.

We ask the probability of spacetime and Langevin time average of the
field variables taking a specific value,
\begin{align}
&\text{Prob}\left(\frac{1}{N^dN_\tau}\sum_n\sum_{\kappa=1}^{N_\tau}\hat{\phi}_n(\kappa) = \bar{\phi}\right)  \nonumber \\
&\,=\int[d\hat{\phi}] \, T\left(\hat{\phi}(N_\tau)|\hat{\phi}(0)\right)\delta\left(\sum_n\sum_{\kappa=1}^{N_\tau}\hat{\phi}_n^R(\kappa) - N^dN_\tau\bar{\phi}^R\right)\delta\left(\sum_n\sum_{\kappa=1}^{N_\tau}\hat{\phi}_n^I(\kappa) - N^dN_\tau\bar{\phi}^I\right) \nonumber \\
&\, = \int [d\hat{\phi}] \, T\left(\hat{\phi}(N_\tau)|\hat{\phi}(0)\right) \int d\lambda^R d\lambda^I\exp\left[i2\pi\lambda^R\left(\sum_n\sum_{\kappa=1}^{N_\tau}\hat{\phi}_n^R(\kappa) - N^dN_\tau\bar{\phi}^R\right)\right]\nonumber \\
&\qquad\qquad\qquad\qquad\times\exp\left[i2\pi\lambda^I\left(\sum_n\sum_{\kappa=1}^{N_\tau}\hat{\phi}_n^I(\kappa) - N^dN_\tau\bar{\phi}^I\right)\right] \nonumber \\
&\, = \int d\lambda^Rd\lambda^I\, Z_d[\hat{J}^R=i2\pi\lambda^R,\hat{J}^R=i2\pi\lambda^I]\exp\left[-i2\pi N^dN_\tau\left(\lambda^R\bar{\phi}^R+\lambda^I\bar{\phi}^I\right)\right].
\end{align}
At this point we will make an approximation, though we will argue that our
approximation becomes exact in the infinite Langevin time limit. We are
looking for a simplification of the effective action evaluated in the presence
of a constant source field. For this purpose, we consider a derivative
expansion of the effective action. The first term will be the effective
potential, as defined above. Other terms will have factors of (lattice)
derivatives of the field variable $\hat\phi_n(\kappa)$. Note that we are
now regarding the field variable as the expactation value of a ``quantum''
operator of the lattice field theory given by \eqref{Zd}. Consider complex
Langevin initial conditions that have no space-time dependence. Then, for a
constant source field, one expects the derivatives along the space-time
directions to vanish due to periodic boundary conditions and translation
invariance. On the Langevin time direction translation invariance breaks,
because of the existence of an initial condition. Even though this is the
case, we will argue that in the long Langevin time limit, one can ignore
higher order terms and safely approximate the effective action by the
effective potential. Consider for a moment a complex Langevin simulation
starting with space-time independent initial conditions. Assuming convergence
to a stationary distribution, in the long Langevin time limit the field
expectation values will ``forget'' the initials conditions and approach a
stationary value. This means that the Langevin time derivatives of the field
expectation values will be small, in fact will vanish in the long time
limit. By definition, one would be able to calculate this effect using the
generating functional given by \eqref{Zd} by setting the source fields to
zero. We are interested in constant source fields that are not vanishing. Now
let's go back to the Euclidean lattice field theory picture defined by
\eqref{Zd}. For zero source fields, in the very large $\kappa$ region the
lattice looks pretty much translationally invariant as argued above. Adding a
constant source field will not break this invariance. Therefore, vanishing of
the Langevin time derivatives must still hold. We argue that if this is the
case, the effective potential will dominate over all higher order terms in the
derivative expansion. To motivate this last comment a little more rigorously,
consider a constant $\gamma$ (like the effective potential) and a function
$\beta(\kappa)$ (like the higher order terms in the derivative expansion) that
dies as $\kappa\rightarrow\infty$. Now consider the quantity
\begin{align}\label{limit}
\left|\lim_{N_\tau\rightarrow\infty} \frac{\sum_{\kappa=1}^{N_\tau} \beta(\kappa)}{\sum_{\kappa=1}^{N_\tau} \gamma}\right|. 
\end{align}
We will show that this limit is zero, which will prove that our comment is valid, that the effective potential dominates over higher terms in the derivative expansion. To prove that it is zero, we will show that it is smaller than any positive $\xi$. We will need to recall the definition of a convergent sequence. A sequence $Q_\kappa$ converges to the limit $Q$ if, for any $\epsilon > 0$, there exists a $P$ such that $\left|Q_\kappa-Q\right|<\epsilon$ for $\kappa>P$. Now we apply this definition to sequence $\beta(\kappa)$, such that $\epsilon = \left|\gamma\right|\xi$. Now there exists a $P_\xi$ such that $\left|\beta(\kappa)\right|<\left|\gamma\right|\xi$ for $\kappa>P_\xi$. Then
\begin{align}
\left|\lim_{N_\tau\rightarrow\infty} \frac{\sum_{\kappa=1}^{N_\tau} \beta(\kappa)}{\sum_{\kappa=1}^{N_\tau} \gamma}\right| &= \left|\lim_{N_\tau\rightarrow\infty} \left(\frac{1}{N_\tau}\sum_{\kappa=1}^{P_\xi}\frac{\beta(\kappa)}{\gamma}+\frac{1}{N_\tau}\sum_{\kappa=P_\xi+1}^{N_\tau}\frac{\beta(\kappa)}{\gamma}\right)\right|=\left|\lim_{N_\tau\rightarrow\infty}\frac{1}{N_\tau}\sum_{\kappa=P_\xi+1}^{N_\tau}\frac{\beta(\kappa)}{\gamma}\right| \nonumber \\
& \leq \lim_{N_\tau\rightarrow\infty} \frac{1}{N_\tau}\sum_{\kappa=P_\xi+1}^{N_\tau}\left|\frac{\beta(\kappa)}{\gamma}\right| < \lim_{N_\tau\rightarrow\infty} \frac{1}{N_\tau}\sum_{\kappa=P_\xi+1}^{N_\tau}\left|\frac{\gamma\xi}{\gamma}\right| = \xi \nonumber \\
\implies
&\left|\lim_{N_\tau\rightarrow\infty} \frac{\sum_{\kappa=1}^{N_\tau} \beta(\kappa)}{\sum_{\kappa=1}^{N_\tau} \gamma}\right| < \xi \qquad \text{for any $\xi>0$}.
\end{align}
Therefore, we can safely assume that a constant source field implies constant $\hat{\phi}_n^J(\kappa)$. As discussed above, this constant $\hat{\phi}_n^J$ is the expectation value of the field variable in the long Langevin time limit in the presence of a constant source field. Then
\begin{align}
   &Z_d[\hat{J}^R=i2\pi\lambda^R,\hat{J}^R=i2\pi\lambda^I] = \exp\left\lbrace N^dN_\tau\left(i2\pi\lambda^R\hat{\phi}^{R,\lambda} + i2\pi\lambda^I\hat{\phi}^{I,\lambda} - \mathcal{V}[\hat{\phi}^{R,\lambda},\hat{\phi}^{I,\lambda}] \right)\right\rbrace.
\end{align}
Here the superscpript $\lambda$ is used to denote that fields are functions of the constant source field. We will look at infinite Langevin time limit. Then using the method of stationary phase
\begin{align}
  \text{Prob}\left(\frac{1}{N^dN_\tau}\sum_n\sum_{\kappa=1}^{N_\tau}\hat{\phi}_n(\kappa) = \bar{\phi}\right) \propto \exp\left[-N^dN_\tau \mathcal{V}(\bar{\phi^R},\bar{\phi^I})+\mathcal{O}\left(1\right)\right]. 
\end{align}
Thus, the effective potential governs the probability distribution of the spacetime and Langevin time average of the field. As the simulation time $N_\tau$ goes to infinity, the space-time and Langevin time average of the fields must equal to the minimum of the effective potential\footnote{We assume no degeneracy among the minima.}. Our argument for effective potential dominance assumed space-time indenpendent initial conditions. However, the stationary distribution that the complex Langevin equation converges should be independent of the initial condition. Therefore this result can be generalized to any initial conditions. Note that if our argument on the dominance of the effective potential is valid, this is an exact result in the infinite $N_\tau$ limit.

This interpretation of the effective potential will enable us to determine the
stationary distribution to which the complex Langevin equation converges.

\subsection{Effective Potential in One Loop}

In this section, we will calculate effective potential to first order in $\Omega$ in the infinite Langevin time limit. As we showed in the previous section, in this limit the space-time and Langevin time average of the field must equal to the minimum of the effective potential. 

Calculation of the effective potential for a complex Langevin simulation that runs for a finite time is not easy. The calculation requires, as we will see, the evaluation of a $2N_\tau N^d$-by-$2N_\tau N^d$ matrix determinant and to the authors, there does not seem to be a way of simplifying this calculation. Instead we will do our calculation in the infinite Langevin time limit, but we will take this limit in a different way than what we have been considering until now. Until now, we let the complex Langevin simulation to start at $\kappa=0$ with a fixed initial condition and run until a finite time $N_\tau$. Then we took the limit $N_\tau\rightarrow\infty$. Instead, we now start our simulation at time $\kappa=-N_\tau$ with a fixed initial condition and run it until $N_\tau$. Then we let $N_\tau$ go to infinity. In the large $N_\tau$ limit, due to ergodicity and Langevin time independence of the complex Langevin equation, the Langevin time averages of these two processes will converge to the same stationary distribution. Another way to see the equivalence is by shifting the latter process by $N_\tau$ steps in Langevin time and thus reaching a process of the former type. Note that this changing of limiting process does not effect the Schwinger-Dyson equation related arguments. Since stationary distributions sampled by the complex Langevin equation for both processes are the same, we can calculate the effective potential for the latter process, and its minimum will match the minimum of the former one. Matching of minima is sufficient for our purposes, but it is very likely true that the effective actions themselves match also. From now on, we work in the limit $N_\tau\rightarrow\infty$ where the limit is taken in the latter way.    
 
We start by expanding the integrand of the generating functional for dynamical correlation functions around constant field configurations $\bar{\phi^R}$ and $\bar{\phi^I}$. First, we choose the constant source fields as 
\begin{align}
\bar{J}^{R,I} &= \frac{1}{\Omega}\left.\frac{\partial I[\hat{\phi}^R,\hat{\phi}^I]}{\partial \hat{\phi}_n^{R,I}(\kappa)}\right|_{\bar{\phi}}.
\end{align}
Using these values of the source fields, we get 
\begin{align}\label{M}
&\exp{W[\bar{J}^R,\bar{J}^I]} =\exp{\left[\sum_{n}\sum_{\kappa}\left(\bar{J}^R\bar{\phi^R}+\bar{J}^I\bar{\phi}^I\right)\right]} \nonumber \\
&\qquad\times\exp {\left[-\frac{\hat{\epsilon}}{4\Omega}\sum_n\sum{\kappa}\left[\frac{\left[\hat{F}^R(\bar{\phi}^R,\bar{\phi}^I)\right]^2}{\alpha}+\frac{\left[\hat{F}^I(\bar{\phi^R},\bar{\phi}^I)\right]^2}{\alpha-1}\right]\right]} \nonumber \\ 
&\qquad\times \mathcal{C} \int [d\hat{\phi}]\, \exp\left[-\frac{1}{2\Omega}\sum_{n,m}\sum_{\kappa,\gamma} \sum_{A,B = R,I}\hat{\phi}_n^A(\kappa) M_{n\kappa A,m\gamma B}\hat{\phi}_m^B(\gamma)+ \mathcal{O}(\hat{\phi}^3)\right]
\end{align}
Here the $\kappa$ sums go from $-\infty$ to $\infty$. The matrix $M$ is given by
\begin{align}\label{Mel}
  &M_{n \kappa R,m\gamma R} = \left.\frac{1}{2\hat{\epsilon}\alpha}\sum_{p\sigma}\hat{\mathcal{D}}_{p\sigma,n\kappa}\hat{\mathcal{D}}_{p\sigma,m\gamma} +  \frac{1}{2\alpha}\left(\hat{\mathcal{D}}_{n\kappa,m\gamma} + \hat{\mathcal{D}}_{m\gamma,n\kappa}\right)\frac{\partial \hat{F}^R}{\partial \hat{\phi}^R}\right|_{\bar{\phi}^R,\bar{\phi}^I} \nonumber \\ 
&\,\left.+\frac{\delta_{nm}\delta_{\gamma\kappa}\hat{\epsilon}}{2}\left[\frac {1}\alpha \left(\frac{\partial \hat{F}^R}{\partial \hat{\phi}^R}\right)^2+\frac {1}\alpha \hat{F}^R\frac{\partial^2 \hat{F}^R}{\partial \hat{\phi}^{R\,2}}+\frac{1}{\alpha-1}\left(\frac{\partial \hat{F}^I}{\partial \hat{\phi}^R}\right)^2 +\frac{1}{\alpha-1}\hat{F}^I\frac{\partial^2 \hat{F}^I}{\partial \phi^{R\,2}}\right]\right|_{\bar{\phi}^R,\bar{\phi}^I},\nonumber \\
&M_{n\kappa I, m\gamma I} = \left.\frac{1}{2\hat{\epsilon}(\alpha-1)}\sum_{p\sigma}\hat{\mathcal{D}}_{p\sigma,n\kappa}\hat{\mathcal{D}}_{p\sigma,m\gamma} +  \frac{1}{2(\alpha-1)}\left(\hat{\mathcal{D}}_{n\kappa,m\gamma} + \hat{\mathcal{D}}_{m\gamma,n\kappa}\right)\frac{\partial \hat{F}^I}{\partial \hat{\phi}^I}\right|_{\bar{\phi}^R,\bar{\phi}^I} \nonumber \\ 
&\,\left.+\frac{\delta_{nm}\delta_{\gamma\kappa}\hat{\epsilon}}{2}\left[\frac {1}{(\alpha-1)} \left(\frac{\partial \hat{F}^I}{\partial \hat{\phi}^I}\right)^2+\frac {1}{\alpha-1} \hat{F}^I\frac{\partial^2 \hat{F}^I}{\partial \hat{\phi}^{I\,2}}+\frac{1}{\alpha}\left(\frac{\partial \hat{F}^R}{\partial \hat{\phi}^I}\right)^2 +\frac{1}{\alpha}\hat{F}^R\frac{\partial^2 \hat{F}^R}{\partial \phi^{I\,2}}\right]\right|_{\bar{\phi}^R,\bar{\phi}^I},\nonumber \\ 
&M_{n\kappa I,m\gamma R}=M_{m\gamma R,n\kappa I} = \left.\frac{1}{2}\left(\frac{\hat{\mathcal{D}}_{n\kappa,m\gamma}}{\alpha} - \frac{\hat{\mathcal{D}}_{m\gamma,n\kappa}}{\alpha-1}\right) \frac{\partial \hat{F}^R}{\partial \hat{\phi}^I}\right|_{\bar{\phi}^R\bar{\phi}^I} \nonumber \\
&\,+\left.\frac {\delta_{nm}\delta_{\kappa\gamma}\hat{\epsilon}}{2}\left[\frac{1}{\alpha}\hat{F}^R\frac{\partial^2 \hat{F}^R}{\partial \hat{\phi}^R\partial \hat{\phi}^I}+\frac{1}{\alpha-1}\hat{F}^I\frac{\partial^2 \hat{F}^I}{\partial \hat{\phi}^R\partial \hat{\phi}^I}+\frac{1}{\alpha(1-\alpha)}\frac{\partial \hat{F}^R}{\partial \hat{\phi}^R}\frac{\partial \hat{F}^R}{\partial \hat{\phi}^I} \right]\right|_{\bar{\phi}^R,\bar{\phi}^I},
\end{align}
where 
\begin{align}
\hat{\mathcal{D}}_{n\kappa,m\gamma} = \delta_{nm}\left(\delta_{\kappa+1,\gamma}-\delta_{\kappa\gamma}\right)-\hat{\epsilon}\delta_{\kappa\gamma}\sum_k\left(\delta_{n+\hat{e}_k,m}+\delta_{n-\hat{e}_k,m} - 2\delta_{nm}\right).
\end{align}
When deriving these equations, we made use of the Cauchy-Riemann equations. They are evaluated at the constant field configurations $\bar{\phi}^R$ and $\bar{\phi}^I$.  Now, using the definition of the effective potential, it is easy to see that
\begin{align}\label{efac}
  \mathcal{V}(\bar{\phi^R},\bar{\phi^I}) &= \frac{1}{\Omega}\frac{\hat{\epsilon}\left[\hat{F}^R\left(\bar{\phi}^R,\bar{\phi}^I\right)\right]^2}{4\alpha} +  \frac{1}{\Omega}\frac{\hat{\epsilon}\left[\hat{F}^I\left(\bar{\phi}^R,\bar{\phi}^I\right)\right]^2}{4(\alpha-1)} + \frac{\ln \det M}{2\sum_n\sum_{\kappa}}  + \mathcal{O}(\Omega).
\end{align}
%

Before continuing our calculation, we stop to discuss this
result. As expected, the leading order term sets the minima of the effective
potential at the classical fixed points, which are saddle points of the action
$S$. This is just saying that if there were no noise (and the simulation did
not diverge), the fields will evolve to a fixed point and stay there for
ever. Averages over space-time and Langevin time in the infinite Langevin time
limit will be set by the classical fixed points. At the classical level, all
saddle points of $S$ are degenerate, meaning they are all minima of the
effective potential. Adding noise will remove the degeneracy through higher
terms in the $\Omega$ expansion. We will calculate the determinant term below
and show in a specific example how it removes the degeneracy.

On a different line of argument, we showed above that if the complex Langevin
equation converges to a stationary distribution, this distribution will be a
solution of the Schwinger-Dyson equation for the theory with complex action
$S$. Moreover, we showed that $\Omega$ counts the number of derivatives
(loops) for this theory. Now we merge this result with the discussion of the
previous paragraph as follows:

\textit{Assuming ergodicity, averages over the stationary distribution is
  equivalent to averages over Langevin time in the infinite Langevin time
  limit. Since we set periodic boundary conditions on the space-time lattice,
  the resulting stationary distribution will have translation invariance in
  space-time. In particular, the Langevin time averages for the field variable
  $\hat{\phi}$ will be constant in space-time. Therefore, we can say that
  Langevin time average of the field variable $\hat{\phi}$ is equal to the
  Langevin time and space-time average of $\hat{\phi}$. Based on our results
  above, we know that the latter is equal to the minimum of the effective
  action for the complex Langevin equation which is given in equation
  \eqref{efac} to $0^\text{th}$ order in $\Omega$. Since $\Omega$
  simultaneously counts loop order for the stationary distribution, which is a
  solution to the Schwinger-Dyson equations, we conclude that the complex
  Langevin equation converges to a stationary distribution that is quantized
  around a saddle point of the complex action $S$. This saddle point is given
  by the minimum of the effective action of the complex Langevin equation.}

The last statement needs more explanation. First of all, by quantization of a
complex action $S$ we mean not the Feynman path integral quantization but a
Schwinger action principle quantization. From Schwinger action principle, one
derives Schwinger-Dyson equations, which have multiple solutions
\cite{Guralnik07,Garcia96}.\footnote{Other differential equations also follow
  from Schwinger action principle, but it was shown in \cite{Guralnik07} that
  a study of Schwinger-Dyson equations is sufficient since in the continuum
  limit solutions of Schwinger-Dyson equations satisfy the remaining
  differential equations.} A solution of Schwinger-Dyson equations defines a
generating functional for a quantization of $S$. This is consistent with
equation \eqref{st}, where we showed that in zero-dimensions solutions of
Schwinger-Dyson equations can be represented by linear combinations of
complexified path integrals. This result trivially generalizes to a lattice
\cite{Guralnik07, Pehlevan07}.\footnote{To our knowledge generalizations of
  generating functionals with complex integration paths are not yet constructed
  on the continuum. Complex Langevin equations can be of help in defining
  these continuum generating functionals.} Complex Langevin simulations have
only one stationary distribution, the question is to find which quantization
of $S$ this distribution gives. Our argument shows that the stationary
distribution will contain contributions only from one of the saddle points of
$S$.\footnote{Here we assumed that the $0^\text{th}$ order term of the
  equation \eqref{efac} breaks the degeneracy among the minima completely. If
  not, one may need to go to higher orders. If the degeneracy is not broken at
  all, it may be that the stationary distribution contains contributions from
  all degenerate minima. The generating functional for this distribution may
  be defined by a linear combination of the generating functionals that
  contain only contributions from one of the degenerate minima. See
  \cite{Guralnik07}.} One way of making this statement precise is through the
results of \cite{Guralnik07}. There it was shown that for zero-dimensional
theories, loop expansions of Schwinger-Dyson equations can be Borel resummed
to give path integrals of type \eqref{ansatz} with the path $\Gamma$ being
equivalent to a steepest descent path passing through a dominant saddle point
of $S$. It is these generating functionals for zero-dimensions, and their
generalizations to higher dimensions that we argue to be the stationary
distributions of the complex Langevin equations. Examples of this will be
given in the next section.

Now we go back to the determinant calculation. Calculation of the determinant
of matrix $M$ is most easily done in the Fourier basis. The discrete time
Fourier transform and the inverse transform are given by
\begin{align}
\tilde{\phi}_l^{R,I}(\omega) &= \frac{1}{(2\pi)^{(d+1)/2}}\sum_{n} \sum_{\kappa}\hat{\phi}_n^{R,I}(\kappa)\,e^{-i\omega\kappa -in\cdot l}, \nonumber \\
\hat{\phi}_n^{R,I}(\kappa) &= \frac{1}{(2\pi)^{(d+1)/2}}\int_{-\pi}^{\pi}dw \int_{-\pi}^{\pi}dl_1\ldots \int_{-\pi}^{\pi}dl_d \, \tilde{\phi}_l^{R,I}(\omega)\,e^{i\omega\kappa +in\cdot l}.
\end{align}
The nonzero elements of the matrix $M$ in this new basis are:\footnote{recall
the Fourier series representation of the $\delta$ function:
\begin{align}
\delta(\omega)=\frac 1{2\pi}\sum_\kappa e^{-i\kappa \omega}.
\end{align}
}
\begin{align}
  M_{l \omega R, l' \omega' R} = &\delta(l-l')\, \delta(w-w')\left\lbrace\left.\frac{1}{2\hat{\epsilon}\alpha}\left(\tilde{\mathcal{D}}_{l\omega}+\hat{\epsilon}\frac{\partial \hat{F}^R}{\partial \hat{\phi}^R}\right)\left(\tilde{\mathcal{D}}_{l\omega}^*+\hat{\epsilon}\frac{\partial \hat{F}^R}{\partial \hat{\phi}^R}\right)\right|_{\bar{\phi}^R,\bar{\phi}^I}\right. \nonumber \\ 
&\left.\left.+\frac{\hat{\epsilon}}{2}\left[\frac {1}\alpha \hat{F}^R\frac{\partial^2 \hat{F}^R}{\partial \hat{\phi}^{R\,2}}+\frac{1}{\alpha-1}\left(\frac{\partial \hat{F}^I}{\partial \hat{\phi}^R}\right)^2 +\frac{1}{\alpha-1}\hat{F}^I\frac{\partial^2 \hat{F}^I}{\partial \phi^{R\,2}}\right]\right|_{\bar{\phi}^R,\bar{\phi}^I}\right\rbrace,\nonumber \\
M_{l\omega I, l'\omega' I} = &\delta(l-l')\,\delta(w-w')\left\lbrace\left.\frac{1}{2\hat{\epsilon}(\alpha-1)}\left(\tilde{\mathcal{D}}_{l\omega}+\hat{\epsilon}\frac{\partial \hat{F}^I}{\partial \hat{\phi}^I}\right)\left(\tilde{\mathcal{D}}_{l\omega}^*+\hat{\epsilon}\frac{\partial \hat{F}^I}{\partial \hat{\phi}^I}\right)\right|_{\bar{\phi}^R,\bar{\phi}^I}\right. \nonumber \\ 
&\left.\left.+\frac{\hat{\epsilon}}{2}\left[\frac {1}{\alpha-1} \hat{F}^I\frac{\partial^2 \hat{F}^I}{\partial \hat{\phi}^{I\,2}}+\frac{1}{\alpha}\left(\frac{\partial \hat{F}^R}{\partial \hat{\phi}^I}\right)^2 +\frac{1}{\alpha}\hat{F}^R\frac{\partial^2 \hat{F}^R}{\partial \phi^{I\,2}}\right]\right|_{\bar{\phi}^R,\bar{\phi}^I}\right\rbrace,\nonumber \\ 
M_{l\omega I,l'\omega' R} &=M_{l'\omega' R,l\omega I}^* = \delta(l-l')\,\delta(w-w') \left\lbrace\left.\frac{1}{2}\frac{\partial \hat{F}^R}{\partial \hat{\phi}^I}\left(\frac{\tilde{\mathcal{D}}_{l\omega}}{\alpha}-\frac{\tilde{\mathcal{D}}_{l\omega}^*}{\alpha-1}\right)\right|_{\bar{\phi}^R\bar{\phi}^I}\right. \nonumber \\
&+\left.\left.\frac {\hat{\epsilon}}{2}\left[\frac{1}{\alpha}\hat{F}^R\frac{\partial^2 \hat{F}^R}{\partial \hat{\phi}^R\partial \hat{\phi}^I}+\frac{1}{\alpha-1}\hat{F}^I\frac{\partial^2 \hat{F}^I}{\partial \hat{\phi}^R\partial \hat{\phi}^I}+\frac{1}{\alpha(1-\alpha)}\frac{\partial \hat{F}^R}{\partial \hat{\phi}^R}\frac{\partial \hat{F}^R}{\partial \hat{\phi}^I} \right]\right|_{\bar{\phi}^R,\bar{\phi}^I}\right\rbrace.
\end{align}
Here,
\begin{align}
 \tilde{\mathcal{D}}_{l\omega}= e^{iw}-1 + 4\hat{\epsilon}\sum_{k=1}^d \sin^2 \frac{l_k}{2}.
\end{align}
It is now easy to calculate the logarithm of the determinant of this block-diagonal matrix where each block is a 2-by-2 matrix.
\begin{align}
  \frac{\det M}{2\sum_n\sum_\kappa} = \frac 12 \int_{-\pi}^{\pi}\frac{dw}{2\pi} \int_{-\pi}^{\pi}\frac{dl_1}{2\pi}\ldots \int_{-\pi}^{\pi}\frac{dl_d}{2\pi} \,\ln\left( M_{l \omega R,l \omega R} M_{l \omega I,l \omega I}- {\left|M_{l \omega R,l \omega I}\right|}^2\right).
\end{align}

Using these results, the final expression for effective action in one loop is
\begin{align}\label{final}
\lim_{N_\tau\rightarrow\infty} &\mathcal{V}(\bar{\phi^R},\bar{\phi^I}) = \frac{1}{\Omega}\frac{\hat{\epsilon}\left[\hat{F}^R\left(\bar{\phi}^R,\bar{\phi}^I\right)\right]^2}{4\alpha} +  \frac{1}{\Omega}\frac{\hat{\epsilon}\left[\hat{F}^I\left(\bar{\phi}^R,\bar{\phi}^I\right)\right]^2}{4(\alpha-1)} \nonumber \\
&+  \frac 12 \int_{-\pi}^{\pi}\frac{dw}{2\pi} \int_{-\pi}^{\pi}\frac{dl_1}{2\pi}\ldots \int_{-\pi}^{\pi}\frac{dl_d}{2\pi} \,\ln\left( M_{l \omega R,l \omega R} M_{l \omega I,l \omega I}- {\left|M_{l \omega R,l \omega I}\right|}^2\right) + \mathcal{O}(\Omega).
\end{align}
In the next section, we will put equation \eqref{final} into use for specific examples. 




\section{An Example: Zero-Dimensional Cubic Theory}\label{Examples}

\begin{figure}
\begin{center}
\begin{tabular}{c}
\includegraphics[height=8.0cm, angle=0 ]
 {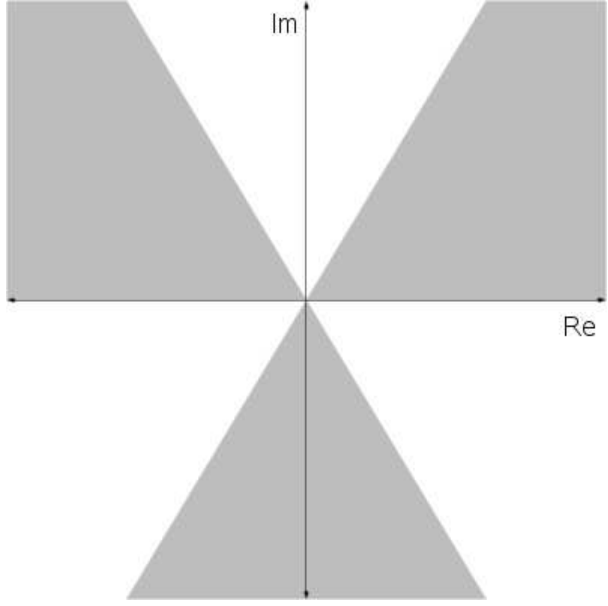}\end{tabular}
\end{center}
\caption{The region defined by $\sin(3\theta)>0$, where $\theta$ is the argument of $\phi$ is shaded. Each wedge is $\pi/3$ radians wide. Any path starting and ending at the infinity within this area corresponds to a particular solution of ultralocal $\phi^3$ theory. 
\label{FigSectorsPhi^3}}
\end{figure}

In this section, we will study zero-dimensional $\phi^3$ theory as an example. The action is given by,
\begin{equation}\label{zde}
  S = -i\phi - \frac{i}{3}\phi^3,
\end{equation}
while the Schwinger-Dyson equation is
\begin{equation}\label{phi3SD}
  -i\frac{d^2Z}{dj^2} - iZ(j) = j \, Z\left(j\right).
\end{equation}
We can construct the path integral solutions to this equation as:
\begin{equation}
	Z(j) =\frac {\int_\Gamma  d\phi \, \exp\left\lbrace i\phi+\frac{i}{3}\phi^3 + j\phi\right\rbrace}{\int_\Gamma  d\phi \, \exp\left\lbrace i\phi+\frac{i}{3}\phi^3\right\rbrace},
\end{equation}
where $\Gamma$ is a path that starts and ends in one of the wedges shown in Figure \ref{FigSectorsPhi^3}.
Since the Schwinger-Dyson equation is second order, only two of these solutions are independent. One needs two boundary/initial conditions to fix a solution. One condition is given by the normalization $Z(0)=1$. Setting another condition will fix the solution completely. 

Now we turn to the complex Langevin dynamics defined by the action \eqref{zde}. The discretized complex Langevin equation is
\begin{align}
 \phi^R(\kappa+1) &= \phi^R(\kappa) - 2\Delta\tau\phi^R(\kappa)\phi^I(\kappa) + \sqrt{\Delta\tau} \,\hat{\eta}^R(\kappa) \nonumber \\
  \phi^I(\kappa+1) &= \phi^I(\kappa) - \Delta\tau\left(-1 -{\phi^R(\kappa)}^2+{\phi^I(\kappa)}^2\right) + \sqrt{\Delta\tau} \,\hat{\eta}^I(\kappa),
\end{align}
Noise correlators are
\begin{align}
  \left\langle \hat{\eta}^R(\kappa) \, \hat{\eta}^R(\kappa') \right\rangle &= 2\Omega\alpha \delta_{\kappa \kappa'} \nonumber \\
\left\langle \hat{\eta}^I(\kappa) \, \hat{\eta}^I(\kappa') \right\rangle &= 2\Omega(\alpha-1)\delta_{\kappa \kappa'}.
\end{align}
Classical fixed points of the complex Langevin equation is given by $\phi=i$ and $\phi=-i$. As mentioned before, both fixed points are minima for the effective potential in the leading order. The next order for $\phi=i$ is given by 
\begin{align}
\frac{1}{2} \ln \frac{1}{4\left(\Delta\tau\right)^2\alpha(\alpha-1)}+ \int_{-\pi}^{\pi}\frac{dw}{2\pi} \,\ln\left[4\left(\Delta\tau\right)^2 + 4(1-2\Delta\tau)\sin^2\frac \omega 2 \right],
\end{align}
and for $\phi=-i$ it is
\begin{align}
\frac{1}{2} \ln \frac{1}{4\left(\Delta\tau\right)^2\alpha(\alpha-1)}+ \int_{-\pi}^{\pi}\frac{dw}{2\pi} \, \ln\left[4\left(\Delta\tau\right)^2 + 4(1+2\Delta\tau)\sin^2 \frac \omega 2 \right].
\end{align}
The integral in the above expression are convergent and $\phi=i$ classical fixed point leads to a smaller term. Based on our discussion in the previous sections, we conclude that the complex langevin equation will converge a stationary distribution that is quantized around $\phi=i$. To be more precise, we mean that the correlators of this theory are given by
\begin{align}
  \left<\phi^k\right>=\frac {\int_\Gamma  d\phi \, \phi^k \exp\left\lbrace i\phi+\frac{i}{3}\phi^3\right\rbrace}{\int_\Gamma  d\phi \, \exp\left\lbrace i\phi+\frac{i}{3}\phi^3\right\rbrace},
\end{align}
where $\Gamma$ is the steepest descent path that passes from $\phi=i$. This steepest descent path is easy to identify is one changes variables to $\phi=-i\psi$. Then
\begin{align}
  \left<\phi^k\right>=\frac {(-i)^k\int_{\Gamma'}  d\psi \, \psi^k \exp\left\lbrace \psi-\frac{1}{3}\psi^3\right\rbrace}{\int_{\Gamma'}  d\psi \, \exp\left\lbrace \psi-\frac{1}{3}\psi^3\right\rbrace},
\end{align}
and now the saddle point that we want to quantize around is at $\psi=-1$. In this form, $\Gamma'$ is recognized as the steepest descent path that defines an Airy function evaulated at 1, i.e. $\text{Ai}(1)$. We refer the reader to \cite{Miller06} and \cite{Bender99} for details of this analysis. In particular,
\begin{align}
  \left<\phi\right> = -i\frac{\text{Ai}'(1)}{\text{Ai}(1)} = 1.176i,
\end{align}
which was calculated by MATLAB's built in Airy functions to 3 decimal places. In this equation, $'$ refers to a derivative with respect to the argument. Now, going back to equation \eqref{phi3SD}, one concludes that
\begin{align}
  \left<\phi^2\right> =-1, \qquad  \left<\phi^3\right> =i-\left<\phi\right> = -0.176i,  \qquad \ldots.
\end{align}
$\left<\phi\right>$ fixes all the other correlators through algebraic relations. This is expected since setting $\left<\phi\right>$ to some number means imposing a second condition, aside from normalization, on the solutions of the Schwinger-Dyson equation \eqref{phi3SD}. This fixes the solution completely. 

We expect a long time complex Langevin simulation to give us these correlators. We did 100 complex Langevin simulations that ran from $\tau=0$ to $\tau=10000$. The step size was $\Delta\tau = 1\times10^{-3}$ and $\alpha$ was chosen to be $1.001$. Each simulation started from a random initial condition. The first three correlators calculated were
\begin{align}
  \left<\phi\right>_{CL}= 0.001(5) + 1.176(1)i, &\qquad \left<\phi^2\right>_{CL}= -0.9999(5) + 0.003(13)i, \nonumber \\ \left<\phi^3\right>_{CL}&= -0.006(21) - 0.176(7)i.
\end{align}
Here the Langevin time averages are averaged over 100 simulations and the error bars stand for the standard deviation over 100 simulations. These numbers are in good agreement with theoretical results, as expected.

\begin{figure}
\begin{center}
\begin{tabular}{c}
\includegraphics[height=8.0cm, angle=0 ]
 {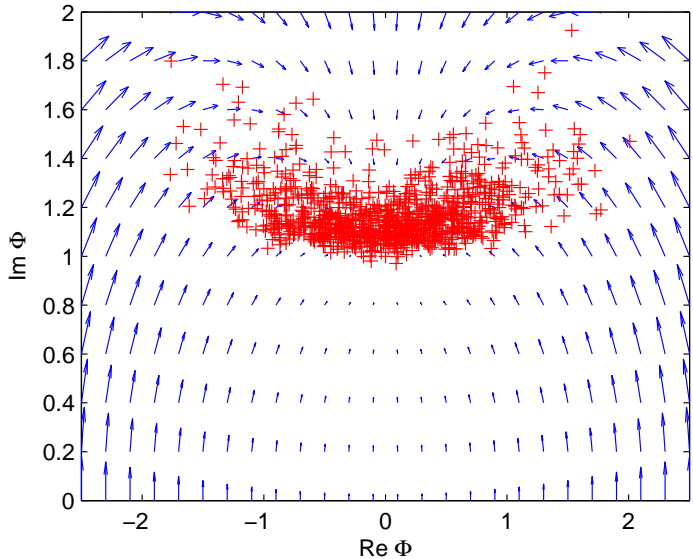}\end{tabular}
\end{center}
\caption{1000 sample points (marked by $+$'s) from a complex Langevin simulation of action \eqref{zde} with classical flow plotted in the background. 
\label{flow}}
\end{figure}

We note that one could expect the $\phi=i$ classical fixed point to dominate over $\phi=-i$ classical fixed point by noticing that the former is a stable fixed point for the classical process, whereas the latter is not. Adding noise eventually makes the simulation sample around the stable fixed point, leading to a quantization around the stable fixed point. This parallels the results of \cite{Aarts08,Berges0708}. Figure \ref{flow} shows a set of sample points from the complex Langevin equation with classical flow on the background. The simulation is seen to sample around the stable fixed point $\phi=i$.
\section{Conclusion and Discussion}

In this paper, we attempted a new approach to the question of identifying the
stationary distribution sampled by a complex Langevin equation. We worked in a
lattice setting. The continuum limit and renormalizability issues are not
trivial and deserves another study of its own.

Our idea is based on a two step process. First, by solving a Schwinger-Dyson
equation \cite{Pehlevan07} one finds a possible set of stationary
distributions. These are complexified path integrals
\cite{Garcia96,Guralnik07}. To identify the particular stationary distribution
sampled by the complex Langevin equation, one needs to use information from
the dynamics of the complex Langevin process. All the information about the
dynamics of a complex Langevin process is in its generating functional. Our
second step, whose details are given in this paper, is to use this generating
functional and define an effective potential based on it. We argued that this
effective potential governs the probability distribution of space-time and
Langevin time averages of the simulated field variable. In particular, in the
infinite Langevin time limit, the field average must equal to the minimum of
the effective potential. We calculated this effective potential, expanding it
in a parameter that multiplies the noise in the complex Langevin equation. In
the lowest order, the minima are given by the saddle points of the action
$S$. The next order removes the degeneracy. The same parameter is shown to count
loops in a loop expansion of the stationary distributions (that satisfy
Schwinger-Dyson equations). Therefore, we concluded that the complex Langevin
equation must be sampling a stationary distribution that describes a field
theory that is quantized around the minimum of the complex Langevin equation
effective action.

 As a final note, we want to mention that this construction can be used as a tool for studying symmetry breaking effects in stochastic quantization for complex actions \cite{Menezes08}.

\section*{Acknowledgments}
The authors would like to thank D. Obeid and D. Ferrante for useful discussions and conversations and the anonymous referee of \cite{Pehlevan07} for bringing the references \cite{Aarts08,Berges0708} to our attention. This work is supported in part by funds provided by the US Department of Energy (DoE) under DE-FG02-91ER40688-TaskD.

\bibliographystyle{elsart-num}

\begin{thebibliography}{10}
\expandafter\ifx\csname url\endcsname\relax
  \def\url#1{\texttt{#1}}\fi
\expandafter\ifx\csname urlprefix\endcsname\relax\def\urlprefix{URL }\fi

\bibitem{Parisi83}
G.~Parisi, On complex probabilities, Phys. Lett. B 131 (1983) 393--395.

\bibitem{Klauder83}
J.~R. Klauder, Stochastic quantization, in: H.~Mitter, C.~B. Lang (Eds.),
  Recent Developments in High-Energy Physics, Springer-Verlag, Wien, 1983, p.
  251.

\bibitem{Salcedo93}
L.~L. Salcedo, Spurious solutions of the complex langevin equation, Phys. Lett.
  B 305 (1993) 125--130.

\bibitem{Pehlevan07}
G.~Guralnik, C.~Pehlevan, {Complex Langevin Equations and Schwinger-Dyson
  Equations}, Nucl. Phys. B811 (2009) 519--536, arXiv:0710.3756.

\bibitem{Aarts08}
G.~Aarts, I.-O. Stamatescu, Stochastic quantization at chemical potential, JHEP
  09 (2008) 018.

\bibitem{Berges0708}
J.~Berges, D.~Sexty, Real-time gauge theory simulations from stochastic
  quantization with optimized updating, Nucl. Phys. B799 (2008) 306--329.

\bibitem{Berges07}
J.~Berges, S.~Borsanyi, D.~Sexty, I.~O. Stamatescu, Lattice simulations of
  real-time quantum fields, Phys. Rev. D 75~(4) (2007) 045007.

\bibitem{Justin02}
J.~Zinn-Justin, Quantum Field Theory and Critical Phenomena, Fourth Edition,
  Oxford University Press, 2002.

\bibitem{Bender77}
C.~M. Bender, F.~Cooper, G.~S. Guralnik, {Path Integral Formulation of Mean
  Field Perturbation Theory}, Ann. Phys. 109 (1977) 165.

\bibitem{Garcia96}
S.~{Garcia}, G.~{Guralnik}, Z.~{Guralnik}, {Theta Vacua and Boundary Conditions
  of the Schwinger Dyson Equations}, hep-th/9612079.

\bibitem{Kawara90}
H.~{Kawara}, M.~{Namiki}, H.~{Okamoto}, S.~{Tanaka}, {Derivation of a
  Generalized Stochastic Path-Integral Formulation Based on Ito Calculus},
  Progress of Theoretical Physics 84 (1990) 749--766.

\bibitem{Gozzi83}
E.~Gozzi, Functional-integral approach to parisi-wu stochastic quantization:
  Scalar theory, Phys. Rev. D 28~(8) (1983) 1922--1930.

\bibitem{Namiki92}
M.~Namiki, Stochastic Quantization, Springer-Verlag, Berlin Heidelberg, 1992.

\bibitem{Coleman88}
S.~{Coleman}, {Aspects of Symmetry}, Aspects of Symmetry, by Sidney Coleman,
  pp.~416.~ISBN 0521318270.~Cambridge, UK: Cambridge University Press, February
  1988., 1988.

\bibitem{Hochberg99}
D.~Hochberg, C.~Molina-Par\'is, J.~P\'erez-Mercader, M.~Visser, Effective
  action for stochastic partial differential equations, Phys. Rev. E 60~(6)
  (1999) 6343--6360.

\bibitem{Crisanti95}
A. Crisanti, U. M. B. Marconi, Effective action method for the Langevin equation, Phys. Rev. E 51 (5) (1995) 4237--4245 

\bibitem{Guralnik07}
G.~Guralnik, Z.~Guralnik, Complexified path integrals and the phases of quantum
  field theory, arXiv:0710.1256.

\bibitem{Miller06}
P.~D. Miller, Applied Asymptotic Analysis, AMS Bookstore, 2006.

\bibitem{Bender99}
C.~M. Bender, S.~A. Orszag, {Advanced mathematical methods for scientists and
  engineers I: asymptotic methods and perturbation theory.}, Springer-Verlag
  Berlin Heidelberg, 1999.

\bibitem{Menezes08}
G.~Menezes, N.~F. Svaiter, Stochastic quantization for complex actions, Journal
  of Mathematical Physics 49~(10) (2008) 102301.

\end{thebibliography}

\end{document}